\documentclass[12pt]{iopart}

\usepackage{iopams}  
\usepackage{graphicx}
\usepackage[breaklinks=true,colorlinks=true,linkcolor=blue,urlcolor=blue,citecolor=blue]{hyperref}

\begin{document}

\title[]{Tuning the electronic hybridization in the heavy fermion cage compound YbFe$_{2}$Zn$_{20}$ with Cd-doping}

\vspace{2pc}

\author{M. Cabrera-Baez, R. A. Ribeiro and M. A. Avila}

\vspace{2pc}

\address{CCNH, Universidade Federal do ABC (UFABC), Santo Andr\'e, SP, 09210-580 Brazil}

\ead{michael.cabrera@ufabc.edu.br, avila@ufabc.edu.br}

\begin{abstract}

Tuning of the electronic properties of heavy fermion compounds by chemical substitutions provides excellent opportunities to further understand the physics of hybridized ions in crystal lattices. 
Here we present an investigation on the effects of Cd doping in flux-grown single crystals of the complex intermetallic cage compound YbFe$_{2}$Zn$_{20}$, that has been described as a heavy fermion with Sommerfeld coefficient of 535 mJ/mol.K$^{2}$. 
Substitution of Cd for Zn disturbs the system by expanding the unit cell and, in this case, the size of the Zn cages that surround Yb and Fe. 
With increasing amount of Cd, the hybridization between Yb $4f$ electrons and the conduction electrons is weakened, as evidenced by a decrease in the Sommerfeld coefficient, which should be accompanied by a valence shift of the Yb$^{3+}$ due to the negative chemical pressure effect. 
This scenario is also supported by the low temperature dc-magnetic susceptibility, that is gradually suppressed and evidences an increment of the Kondo temperature, based on a shift to higher temperatures of the characteristic broad susceptibility peak. 
Furthermore, the DC resistivity decreases with the isoelectronic Cd substitution for Zn, contrary to the expectation for an increasingly disordered system, and implying that the valence shift is not related to charge carrier doping.
The combined results demonstrate excellent complementarity between positive physical pressure and negative chemical pressure, and point to a rich playground for exploring the physics and chemistry of strongly correlated electron systems in the general family of Zn$_{20}$ compounds, despite their structural complexity.

\end{abstract}


\submitto{\ J. Phys.: Condens. Matter}

\maketitle

\section{Introduction}

Heavy fermion compounds play a central role in strongly correlated electron systems, and their physical properties have been studied extensively over the past three decades.\cite{Stewart, Stewart1, Stewart2, Misra} 
The exotic low temperature phenomena involved in this kind of system are associated with an entropy transfer from local moment degrees of freedom to the conduction electrons ($ce$). 
A defining feature of the heavy fermion compounds is a largely enhanced Sommerfeld coefficient $\gamma$, as exemplified historically by CeCu$_{2}$Si$_{2}$ which exhibits $\gamma\sim1100$~mJ/mol.K$^{2}$, implying a renormalized effective mass of $m^{*}$ $\approx$ 10$^{2}$~$m_{e}$,\cite{Stewart} where $m_{e}$ is the free electron mass.
The level of degeneracy of the electronic levels near the Fermi surface in the involved rare-earth ions also plays a fundamental role in the hybridized $4f-ce$ properties.\cite{Rajan} 

Rare-earth heavy fermion systems\cite{Stewart} are customarily accompanied by the Kondo effect, in which the hybridized moment-bearing ions in the lattice produce a characteristically enhanced scattering of the charge carriers, manifested as a local minimum and logarithmic increase of the resistivity with decreasing temperature.\cite{Kondo} 
Upon further cooling, dense Kondo systems also crossover into a coherent scattering regime and a Fermi-liquid ground state is established.\cite{Lengyel} 
These phenomena modify not only the transport but also the magnetic behavior at low temperatures, suppressing the susceptibility of the high temperature paramagnetic state.\cite{Stewart}
Further manifestations appear in heat capacity measurements from which the estimated Sommerfeld coefficient, that accounts for the total electronic density of states, takes an unusually high value that is associated with the electron effective mass renormalization.

\begin{figure}[!ht]
\begin{center}
\includegraphics[width=85mm,keepaspectratio]{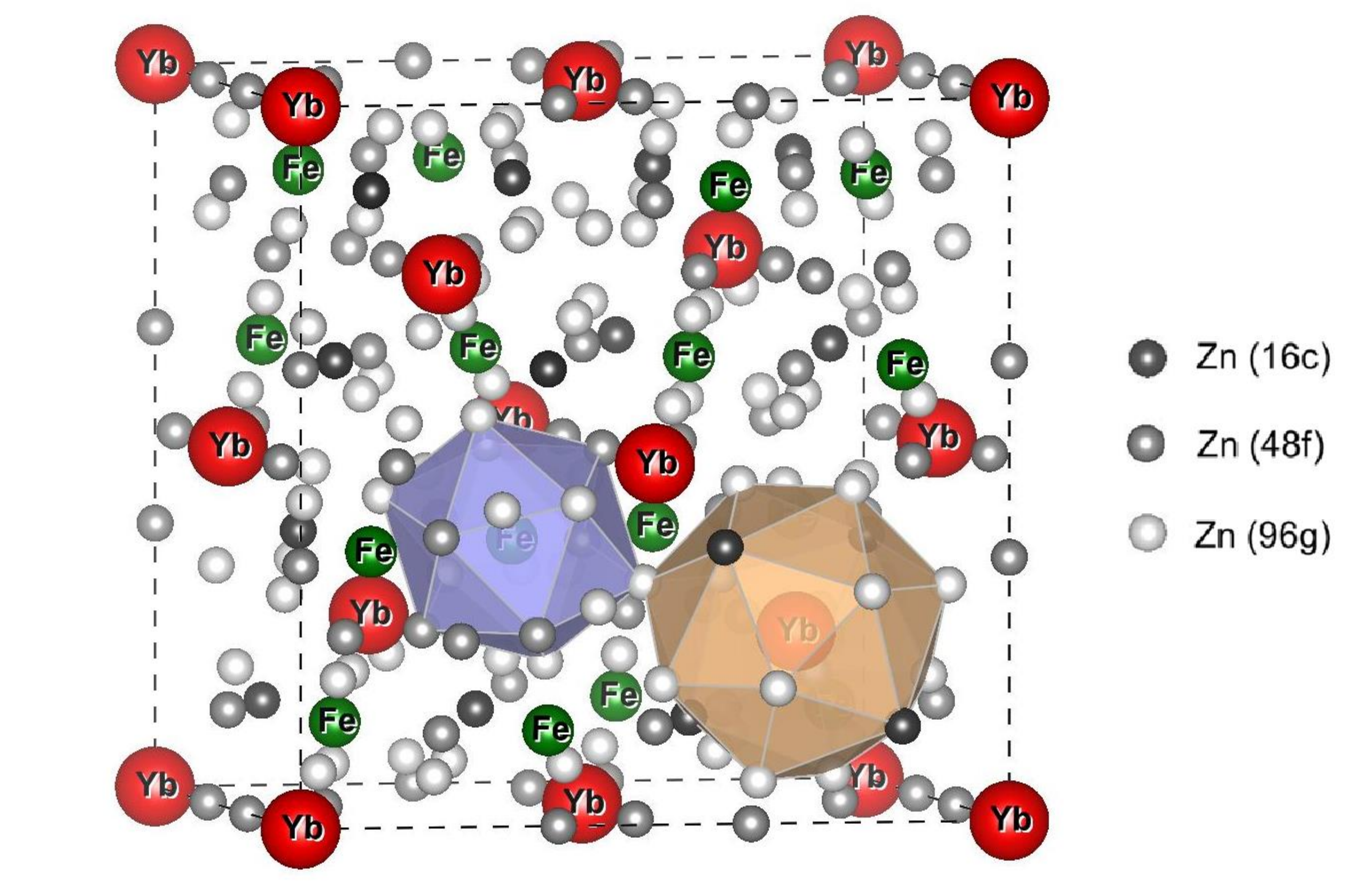}
\end{center}
\vspace{-0.5cm} \caption{Representation of the cubic unit cell of YbFe$_{2}$Zn$_{20}$ ($a \approx 14$~\AA), highlighting the CN-16 Frank-Kasper polyhedron around the rare earth ions and the icosahedron around the transition metal ions.}\label{structure}
\end{figure}

A few years ago, Torikachvili \emph{et al.} characterized six new Yb-based heavy fermion intermetallic compounds\cite{Torikachvili} in the YbT$_{2}$Zn$_{20}$ family (T = Fe, Co, Ru, Rh, Os and Ir) with Sommerfeld coefficients $\gamma$ $>$ 400 mJ/mol K$^{2}$.
The RT$_{2}$Zn$_{20}$ family adopts the complex cubic CeCr$_{2}$Al$_{20}$-type structure with space group $Fd\bar{3}m$.\cite{Nasch}
R and T atoms occupy unique crystallographic sites (Wyckoff positions $8a$ and $16d$, respectively) and the Zn atoms occupy three different crystallographic sites ($96g$, $48f$ and $16c$).
An important feature in this structure is that the R and T atoms are each surrounded by Zn ions as nearest neighbors, in the form of cages (16 and 12 Zn atoms respectively) that leave a shortest R-R spacing of $r \approx 6$~\AA, as exemplified in Fig.~\ref{structure}.

The large cages made of numerous Zn atoms have an effect of isolating the rare earths from each other, and result in only weak crystal electric field (CEF) effects.\cite{Jia}
Studies on compounds of this family with non-hybridizing ions (R = Y, Gd) have demonstrated that the electronic bands near the Fermi level created by the transition metal and the Zn$_{20}$ complex is by its own merit already rich in physical phenomena.
For example, radically different magnetic behaviors are observed in the case of YCo$_{2}$Zn$_{20}$, that shows Pauli-like paramagnetic behavior (dc magnetic susceptibility almost independent of the temperature) compared to YFe$_{2}$Zn$_{20}$, reported as a compound close to the Stoner limit\cite{Jia2} in which the dc magnetic susceptibility has a strong dependence with the temperature. 
These different ``Fermi seas'', in which the $d$-band filling is very important, also drive the general behaviors of the equivalent compounds containing magnetic rare earths.
As a striking example, one can find low-temperature antiferromagnetic order ($T_N\sim5.7$~K) in GdCo$_{2}$Zn$_{20}$ \cite{Jia2} contrasting with high temperature ferromagnetic order ($T_C\sim86$~K) in GdFe$_{2}$Zn$_{20}$.\cite{Jia0}
In the case of GdCo$_{2}$Zn$_{20}$, we have recently reported that the Ruderman-Kittel-Kasuya-Yosida (RKKY) interaction can adequately explain the antiferromagnetic ordering of this compound\cite{Michael15} as is expected in this diluted rare-earth system without electron-electron correlations. 
For GdFe$_{2}$Zn$_{20}$, however, a model that can quantitatively explain the ferromagnetic order with such high $T_{C}$ is still an open question.\cite{Budko, Jia2} 

In the case of the Yb-based compounds,\cite{Canfield1} a strong influence of the environment in which this hybridizing rare earth is embedded is also notable, such as the two heavy fermion compounds YbFe$_{2}$Zn$_{20}$ with $\gamma \approx 520$~mJ/mol~K$^2$ and YbCo$_{2}$Zn$_{20}$ with $\gamma \approx 7900$~mJ/mol~K$^2$.\cite{Torikachvili, Tanaka}

It is well known that the substitution of one element for another in a compound is one of the most versatile and powerful tools for the materials scientist to disturb the system in a controlled manner and subsequently track the evolution of its properties.
One recent example in this family is the substitution of Al for Zn in the GdFe$_{2}$Zn$_{20}$ compound, that can be understood as a combination of positive chemical pressure plus electron doping,\cite{Ni} leading to a drastic reduction of the ferromagnetic behavior.
Chemical substitutions in the Yb-based members have not been reported yet, but hydrostatic pressures up to 8.23~GPa has been applied in the YbFe$_{2}$Zn$_{20}$ compound,\cite{Kim} driving the characteristic Kondo temperature T$_{K}$ to lower temperatures and showing the loss of Fermi liquid behavior.
A critical pressure of 9.8~GPa was inferred in order to reach the quantum critical point in this system, so it is expected to be moving towards Yb$^{3+}$ valence.
Conversely, in the case of the YbCo$_{2}$Zn$_{20}$ compound,\cite{Saiga} it was suggested that the magnetic ordering transition occurs at a lower pressure, around 1 GPa.
In order to attempt negative chemical pressure without electronic doping in this heavy fermion system, we have performed Cd-doping in single crystals of YbFe$_{2}$Zn$_{20}$.
Small amounts of Cd substitution for Zn indeed expand the crystal lattice as evidenced by x-ray diffraction, but more importantly, produce notorious changes in the magnetic, thermodynamic and transport properties, that complement previous studies applying positive (physical) pressure. 
An Yb valence-shift scenario is then suggested as explanation for the observed physical changes.

\section{Experimental details}

Single crystalline samples of YbFe$_{2}$Zn$_{20-x}$Cd$_{x}$ ($0 \leq x \leq 1.4$) were grown out of excess Zn using standard self-flux method.\cite{Canfield,Raquel} 
The constituent elements were 99.99$\%$ Yb (Ames), 99.9$\%$ Co, 99.9$\%$ Fe, 99.9999$\%$ Zn and 99.9999$\%$ Cd (Alfa-Aesar).
Initial ratios of starting elements were $1:2:47-y:y$ ($y=0-6$) for the pseudo-quaternary system Yb:Fe:Zn:Cd.
The elemental mixtures were sealed in an evacuated quartz ampoule and heated in a box furnace.
Crystals were grown by slowly cooling the melt between 1100~$^\circ$C and 600~$^\circ$C over 100~h.
At 600 $^\circ$C the ampoules were removed from the furnace, inverted and placed in a centrifuge to spin off the excess flux.
The separated crystals are typically polyhedral, $\sim3$~mm or larger.

\begin{figure}[!ht]
\begin{center}
\includegraphics[width=85mm,keepaspectratio]{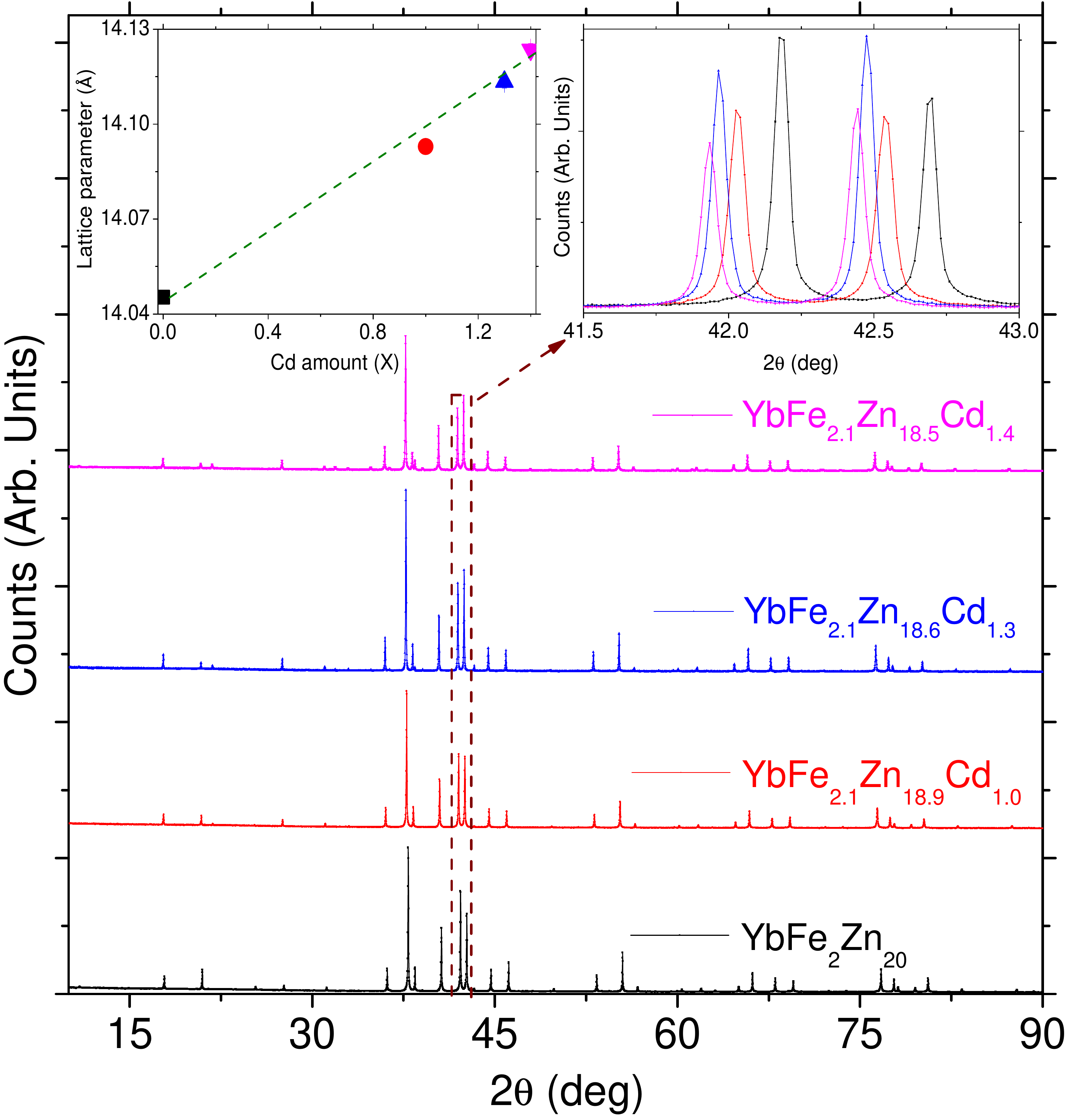}
\end{center}
\vspace{-0.5cm} \caption{Cu-$K_{\alpha}$ powder x-ray diffraction pattern of YbFe$_{2}$Zn$_{20-x}$Cd$_{x}$ ($0 \leq x \leq 1.4$). Left inset shows the evolution of the lattice parameter showing an expansion of the unit cell also evidenced in the right inset that shows a shift to low angles of the diffraction pattern as a function of the Cd amount.}\label{XRD}
\end{figure}

The effective Cd concentrations $x$ on all of our samples were estimated using Energy Dispersive X-ray Spectroscopy (EDS) measurements in a JEOL model JSM-6010LA scanning electron microscope with a Vantage EDS system.
Powder x-ray diffraction on finely crushed crystals at room temperature was used to ascertain the CeCr$_{2}$Al$_{20}$-type structure\cite{Nasch} as exemplified in Fig.~\ref{XRD}. 
Measurements were performed on a STOE STADI-P powder diffractometer in transmission geometry by using a $K_{\alpha 1}$ ($\lambda$ = 1.54056~\AA) wavelength emitted by a Cu anode and selected by a curved Ge(111) crystal, with a tube voltage of 40 kV and a current of 40 mA.
The refined lattice parameter of $a = 14.045(2)$~\AA~ for the YbFe$_{2}$Zn$_{20}$ diffraction pattern is in good agreement with the literature\cite{Jia} and $a = 14.124(2)$~\AA~ for YbFe$_{2}$Zn$_{18.5}$Cd$_{1.4}$ (see Fig.~\ref{XRD}) evidences an expansion of the lattice parameter due to the Cd doping.

Temperature-dependent specific heat ($C_p$) was measured for all of our samples in a Quantum Design Physical Properties Measurement System (PPMS) using the relaxation technique at zero field. 
The DC transport option of the same platform was used to measure temperature-dependent electrical resistivity ($\rho$), through standard four-probe technique in samples that were cut into 3-5 mm bars using a wire saw. 
DC magnetic susceptibility ($\chi=M/H$) measurements were conducted on a Quantum Design MPMS3 magnetometer (SQUID-VSM) under applied fields $H\leq30$~kOe and temperatures in the interval 2.0~K~$\leq T\leq 310$~K.

\section{Experimental Results}

\begin{figure}[!ht]
\begin{center}
\includegraphics[width=85mm,keepaspectratio]{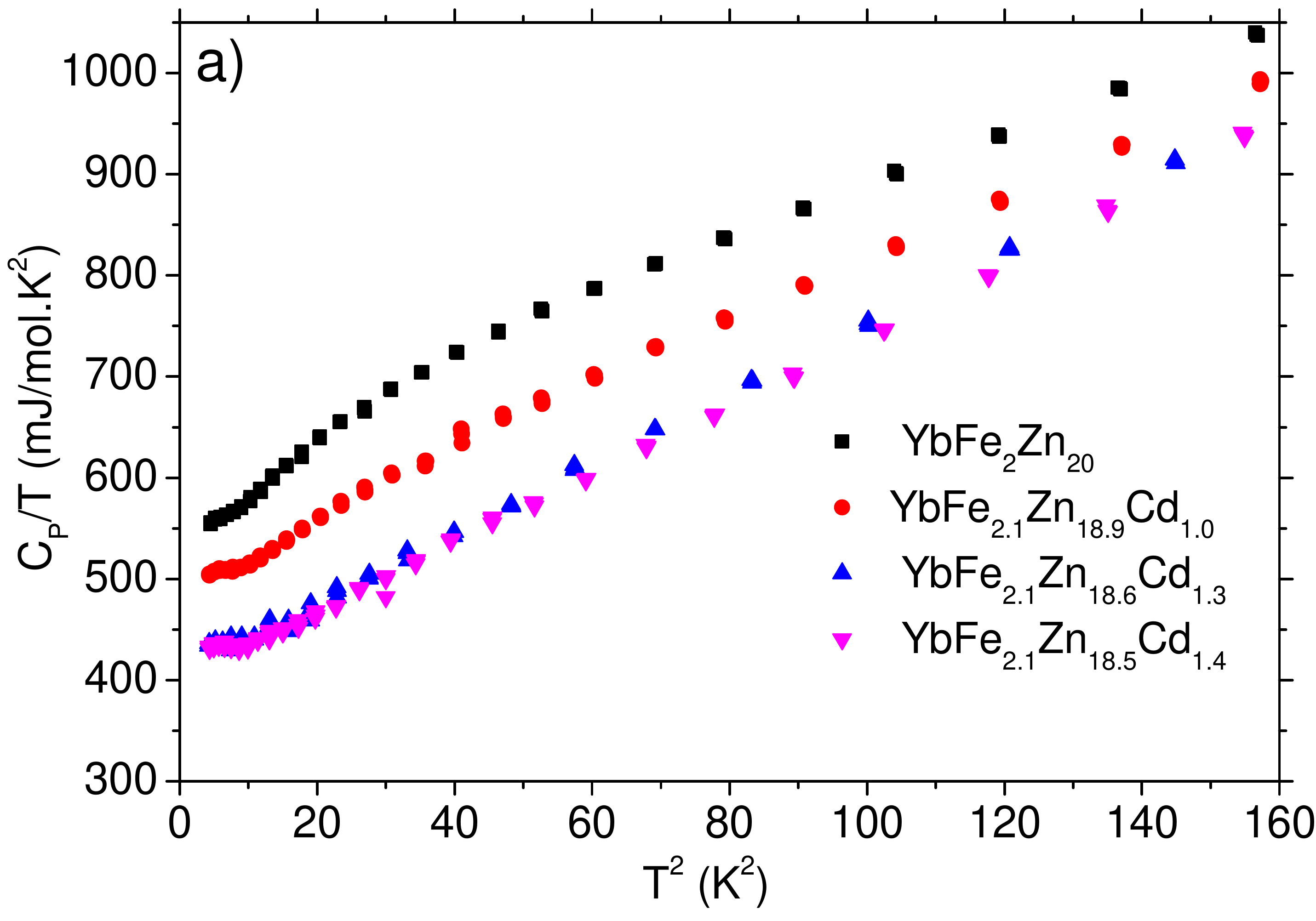}
\includegraphics[width=85mm,keepaspectratio]{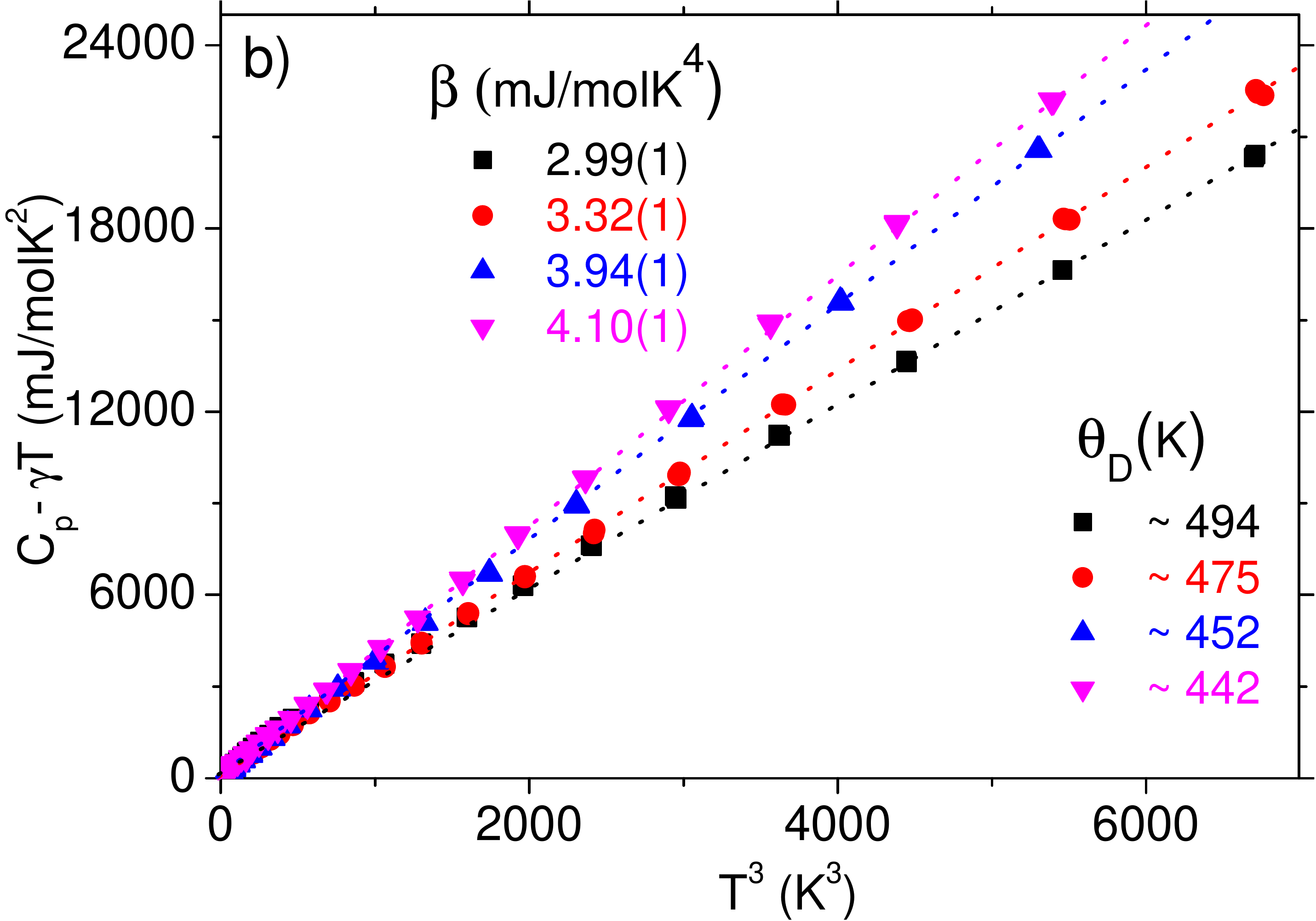}
\end{center}
\vspace{-0.5cm} \caption{(a) Low temperature behaviour of $C_p/T$ as a function of $T^{2}$ for YbFe$_{2}$Zn$_{20-x}$Cd$_{x}$ ($0 \leq x \leq 1.4$). (b) Estimation of the phonon contribution using a plot of $C_{p}-\gamma T$ as a function of $T^{3}$ }
\label{Cp}
\end{figure}

\begin{table}
\centering
\caption{Cd concentrations ($x$), Sommerfeld coefficient $\gamma$, quasiparticle effective mass $m^{*}$, Beta parameter of the specific heat $\beta$, and Debye temperature $\theta_{D}$ for the YbFe$_{2}$Zn$_{20-x}$Cd$_{x}$ system.}
\label{Cp1}
\vspace{+0.5cm}
\begin{tabular}{|c||c||c||c||c|}
 \hline
 Conc.   & $\gamma$    & $m^{*}$ & $\beta$   &   $\theta_{D}$  \\
\hline
\emph{x}  &   (mJ/molK$^{2}$)   & $m_{e}$ & (mJ/molK$^{4}$)  &  (K)\\
  \hline
 0& 535(5) & 764 &2.99(1)  &  494(5) \\
 \hline
 1.0& 498(3) & 712 &3.32(1)   &   475(5)\\
 \hline
 1.3& 430(3) & 614 &3.94(1)  &    452(5) \\
 \hline
 1.4& 425(3)  & 607 &4.10(1)   &   442(5)\\
 \hline
\end{tabular}
\end{table}

Fig.~\ref{Cp}(a) shows the low temperature behavior of $C_p/T$ as a function of $T^{2}$ for YbFe$_{2}$Zn$_{20-x}$Cd$_{x}$, demonstrating a progressive reduction of the Sommerfeld coefficient $\gamma=(C_p/T)|_{T^2\rightarrow0}$, about 100~mJ/mol~K$^2$ lower for $x=1.4$ when compared to the ternary compound.
A limited linear regime gradually extends to lower temperatures in the samples with greater amount of Cd, indicating that the system is evolving towards a better Fermi-Liquid description. 
The Sommerfeld coefficients of the reference compounds YFe$_{2}$Zn$_{20}$ and LuFe$_{2}$Zn$_{20}$ \cite{Jia} ($\gamma$ $\approx$ 50 mJ/mol.K$^{2}$) gives an idea of the Yb effect in this system.
Accordingly, the quasiparticle effective mass decreases from $m^{*}$  $\approx$ 764$m_{e}$ ($x$ = 0) to  $m^{*}$  $\approx$ 607$m_{e}$ ($x$ = 1.4), from a simplified estimate using a parabolic heavy band with $k_{F}$ of conventional metals like Au\cite{Aschcroft} (see Tab.~\ref{Cp1}).
The non-linearities make it difficult to extract the $\beta$ parameter and Debye temperature $\Theta_{D}$ from the standard method of fitting $C_p/T = \gamma +\beta T^2$ in our results. 
In order to obtain a better estimation of the phonon contribution, in Fig.~\ref{Cp}(b) we plot $C_{p}-\gamma T$ as a function of $T^{3}$.
Using $\Theta_{D}=(\frac{12\pi^{4}k_{B}N_{A}Z}{5\beta})^{1/3}$ with $N_{A}=6.02\times10^{23}~$mol$^{-1}$ and $Z=184$ as the number of atoms per unit cell, we obtain the corresponding Debye temperature for each concentration of Cd. 
A gradual decrease in $\Theta_D$ as a function of the Cd amount is evidenced and summarized in Table~\ref{Cp1}.

\begin{figure}[!ht]
\begin{center}
\includegraphics[width=85mm,keepaspectratio]{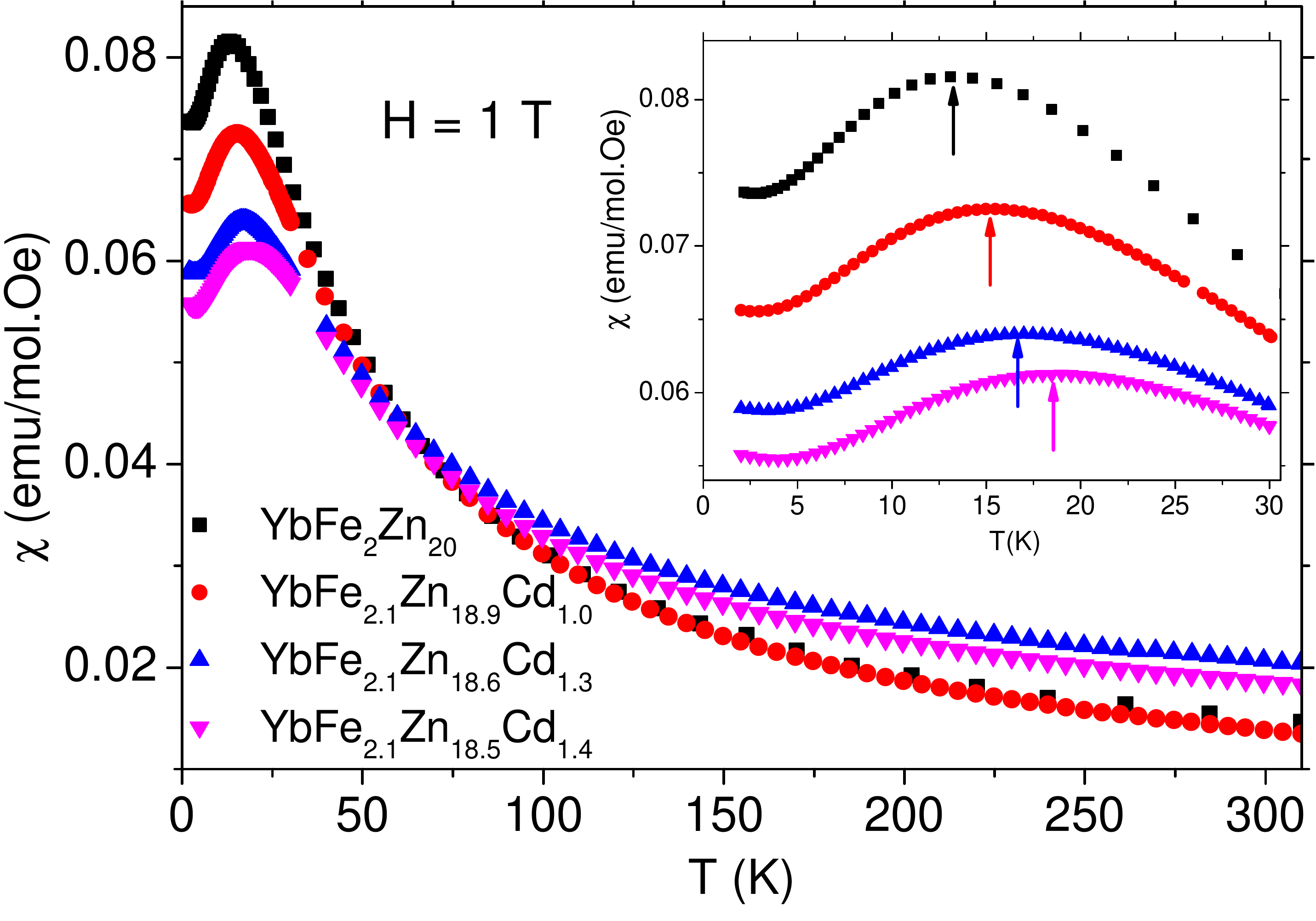}
\end{center}
\vspace{-0.5cm} \caption{$T$-dependence of the dc magnetic susceptibility $\chi$ of YbFe$_{2}$Zn$_{20-x}$Cd$_{x}$ ($0 \leq x \leq 1.4$). The inset shows a zoom in low temperatures showing a shift to high temperatures of the maximum as Cd increases}\label{Suscept}
\end{figure}

Fig.~\ref{Suscept} presents the $T$-dependence of the magnetic susceptibility of YbFe$_{2}$Zn$_{20-x}$Cd$_{x}$ where a maximum is evidenced in all samples, may be related to the screening of the localized moments by the $ce$ or valence-intermediate behavior. This will be discussed ahead.  
There is a progressive reduction of the maximum as a function of Cd amount.
A zoom into the low temperature range is presented in the inset, showing that this decrease is accompanied by a shift to high temperatures of the position of the maximum. 
Extrapolations of $\chi(T)|_{T\rightarrow0}$ also show a reduction as a function of the Cd doping.
Curie-Weiss behavior is observed above 50~K and was fit for all samples (not shown here) giving an increment in the Curie-Weiss temperature $\Theta_{CW}$ with the amount of Cd (from $\sim -17$~K for YbFe$_{2}$Zn$_{20}$ to $\sim -33$~K for the sample with the highest Cd content). 
In contrast, the effective moment per formula unit remains almost unaltered around 5.0~$\mu_B$ within the precision of the analyses. 
It is notably larger than the Yb$^{3+}$ effective moment of 4.54~$\mu_B$ and consistent with a previously reported larger value \cite{Jia} of 4.7~$\mu_B$. 
Note that the reference compound YFe$_{2}$Zn$_{20}$ has been labeled a ``Nearly Ferromagnetic Fermi Liquid'' for being close to the Stoner limit \cite{Jia2} and very recent spectroscopic studies have detected small but discernible moments induced on the cage structure,\cite{Mardegan} which may be involved in the extra contribution to the effective moment.
The obtained results are summarized in the Table~\ref{Suscept1}.

\begin{table}
\centering
\caption{Cd concentrations ($x$), Pauli-like susceptibility ($\chi_{0}$), effective moment ($\mu_{eff}$) and magnetic susceptibility at $T$=0 ($\chi(T=0)$) for the YbFe$_{2}$Zn$_{20-x}$Cd$_{x}$ system.}
\label{Suscept1}
\vspace{+0.5cm}
\begin{tabular}{|c||c||c||c|}
 \hline
 Conc.   & $\chi_{0}$     &   $\mu_{eff}$ &   $\chi(T=0)$ \\
\hline
\emph{x}  &   (emu/mol)   &  ($\mu_{B}$/f.u) &  (emu/mol)\\
  \hline
 0.0& 0.0055(4)      &  4.9(1)      &  0.074(2) \\
 \hline
 1.0& 0.0037(4)   &   5.1(1)     &  0.066(2) \\
 \hline
 1.3& 0.0109(5)    &    5.1(1)    &  0.060(2) \\
 \hline
 1.4& 0.0086(4)    &   5.1(1)     &  0.056(2) \\
 \hline
\end{tabular}
\end{table}

\begin{figure}[!ht]
\begin{center}
\includegraphics[width=85mm,keepaspectratio]{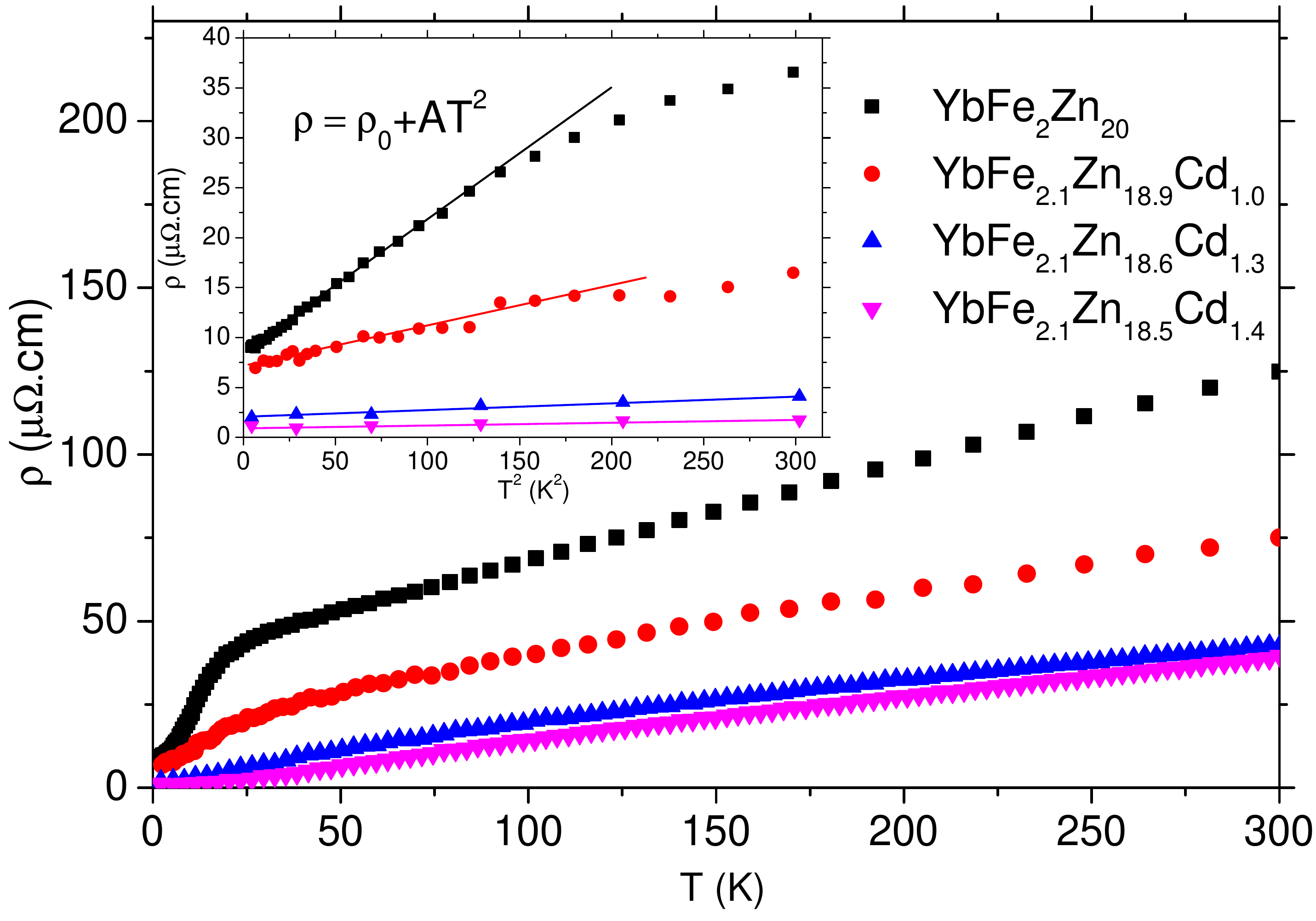}
\end{center}
\vspace{-0.5cm} \caption{Temperature-dependent electrical resistivity ($\rho$) for YbFe$_{2}$Zn$_{20-x}$Cd$_{x}$ ($0 \leq x \leq 1.4$). The inset shows resistivity as a function of $T^2$, highlighting Fermi liquid behavior $\rho$ = $\rho_{0}$ + $AT^{2}$ at the lowest temperatures (solid lines).}\label{Res}
\end{figure}

Fig.~\ref{Res} presents the $T$-dependent electrical resistivity data for our Cd-doped single crystals, which show metallic behavior. 
For the undoped sample a clear coherence shoulder is seen at low temperatures, in agreement with previous studies.\cite{Torikachvili, Kim}
For the doped samples the coherence peak broadens as the overall resistivity decreases strongly in the full range of temperatures. 
The inset shows the low-temperature behavior of the electrical resistivity as a function of $T^{2}$, evidencing Fermi liquid behavior $\rho$ = $\rho_{0}$ + $AT^{2}$ at the lowest temperatures.

\begin{table}
\centering
\caption{Cd concentrations ($x$), residual resistivity ($\rho_{0}$), coefficient of the $T^{2}$ resistivity ($A$) and Wilson Ratio ($R_W$) for the YbFe$_{2}$Zn$_{20-x}$Cd$_{x}$ system.}
\label{tab2}
\vspace{+0.4cm}
\begin{tabular}{|c||c||c||c||c|}
 \hline
 Conc.   & $\rho_{0}$    & $A$$\times$10$^{-1}$      &    $R_W$ \\
\hline
\emph{x}  &   ($\mu \Omega$.cm)  &  ($\frac{\mu \Omega.cm}{K^{2}}$)   &   \\
  \hline
 0                  & 8.4(3)          & 1.30(3)                        &   1.3(1)  \\
 \hline
 1.0               & 7.1(2)          & 0.40(4)                       &   1.1(1)  \\
 \hline
 1.3               & 2.0(2)          & 0.07(1)                         &   1.1(1)  \\
 \hline
 1.4               & 1.0(2)          & 0.02(1)                          &   1.1(1)   \\   
 \hline 

\end{tabular}
\end{table}

\section{Analysis and Discussion}

The combined set of figures and tables presented in the previous section make it evident that, despite the limited amount of Cd ions that could be introduced in the crystals (up to 7\%), the substitutions resulted in significant changes in almost every measured physical property.
Since Cd represents an isoelectronic substitution for Zn, it is reasonable to expect that the primary source of these changes is the crystalline lattice expansion demonstrated in Fig.~\ref{XRD}.

Let us begin by discussing the specific heat results within this framework.
The low-$T$ linear behavior of $C_p/T = \gamma + \beta T^2$ for YbFe$_{2}$Zn$_{20-x}$Cd$_{x}$ of Fig.~\ref{Cp}(a) leads to a strong reduction in the Sommerfeld coefficient $\gamma$ as a function of the Cd content. 
This is then understood as being associated with the reduction of the $f-ce$ hybridization around the Fermi level, that should be caused by a valence-shift of the Yb ion due to the negative chemical pressure effect induced by the Cd ion substitution.
Within the Doniach diagram scenario for Yb-based systems,\cite{Doniach} positive pressures can compress the Yb $4f$ orbitals, moving the ground-state $4f$ energies away from the Fermi level and forcing a shift of the Yb valence towards 3+.
This in turn leads to a more localized character of the $4f$ electrons, a weaker $f-ce$ coupling strength ($J_{eff}$) and eventually to a magnetically ordered ground state governed by the RKKY interaction.
Conversely, ``negative pressures'' as represented by the present Cd doping, allows the $4f^{14}$ state to decrease in energy, towards $4f^{13}$ and through the Fermi level, which increases $J_{eff}$ and favors a Kondo-screened Fermi liquid ground state while hybridization is present, then eventually leads to a non-magnetic (closed-shell) divalent ground state for Yb.
In the whole intermediate valence range, the high degree of $4f-ce$ hybridization results in an enhanced density of states (DOS) around the Fermi level when compared to the end states (Yb$^{3+}$ or Yb$^{2+}$), so enhanced values of $\gamma$ are expected to eventually decrease in either direction towards more typical values of common metals.  

In terms of the phonon contribution to the specific heat, and regarding the Debye temperature $\Theta_D$ as that at which the highest-frequency phonon modes are excited, Fig.~\ref{Cp}(b) implies that Cd substitution for Zn decreases the stiffness of the cage system. 
In principle the Cd atoms may enter in any of the Zn sites ($16c$, $48f$ or $96g$) and there should indeed be a disordered occupation of all three.
However, for LaRu$_{2}$Zn$_{20}$ it was recently reported that the Zn ions at the $16c$ site exhibit low energy, localized vibration modes (\emph{rattling}).\cite{Wakiya}
This is indicative of ``loose'' Zn ions at this site due to the presence of extra space and consequently broadened Madelung potentials.
If so, the larger Cd atoms may be expected to occupy this site preferentially, and the small decrease in the value of $\Theta_D$ is due to the greater atomic mass of Cd.
This preferential occupation of a single site may be partly behind the reasons why our crystal flux growths have shown successful but limited effect in introducing Cd into the YbFe$_{2}$Zn$_{20}$, whose lattice parameter (Fig.~\ref{XRD}) is significantly smaller than $a=14.4263(2)$ of LaRu$_{2}$Zn$_{20}$.
High resolution diffraction experiments will be required to confirm these expectations.

DC magnetic susceptibility measurements (Fig.~\ref{Suscept}) show a broad maximum that is typical of valence fluctuating systems, but can also be related to the Kondo physics \cite{Rivier}.
Well-defined valence fluctuating systems are usually characterized by moderately large values of $\gamma$ [50 - 100 mJ/molK$^{2}$] as in the cases of Yb$_{2}$TGe$_{6}$ \cite{Shigetoh} and Yb$_{2}$Ni$_{12}$As$_{7}$ \cite{Jiang}; Curie-Weiss behavior limited to much higher temperatures (CeSn$_{3}$\cite{Dijkman}, CeRu$_{2}$ \cite{Tsvyashchenko}); effective Yb valence values well shifted away from 3+ (Yb$_{2}$Ni$_{12}$As$_{7}$ \cite{Jiang}, YbNiGe$_{3}$ \cite{Sato}) and a maximum of the magnetic susceptibility at high temperatures associated with the spin-fluctuation temperature $T_{sf}$ \cite{Sales} as in the case of CeRu$_{2}$ \cite{Tsvyashchenko} and Yb$_{2}$Ni$_{12}$As$_{7}$ \cite{Jiang}. 
Conversely, well-defined heavy fermion systems have significantly higher values of $\gamma$ (YbNiSi$_{3}$ \cite{Avila}, YbT$_{2}$Zn$_{20}$ \cite{Torikachvili}), Curie-Weiss behavior extending to lower temperatures (YbNiSi$_{3}$ \cite{Avila}, Yb$_{2}$Ni$_{12}$P$_{7}$ \cite{Jiang}, YbT$_{2}$Zn$_{20}$ \cite{Torikachvili}), Kondo-type interaction (Yb$_{2}$Ni$_{12}$P$_{7}$ \cite{Jiang}) present at low temperatures, and effective Yb valence very close to 3+ (Yb$_{2}$Ni$_{12}$P$_{7}$ \cite{Jiang}, YbNiSi$_{3}$ \cite{Sato, Avila}). 
The crossover region between these two regimes does not have a well-defined limit, so shifts in the maximum of the magnetic susceptibility for such crossover systems can be indicative of shifts in both the Kondo-type interaction ($T_{K}$) and the spin-fluctuation ($T_{sf}$). 
Miccroscopic techniques such as XAS are required in order to quantitatively follow effective Yb valence shifts as a function of Cd doping and temperature.

Fig.~\ref{Suscept} demonstrates the effects of Cd substitution in the system's magnetic response.
Increasing Cd content results in reduction of the magnetic susceptibility at $T$=0 ($\chi(T=0)$).
At high temperatures the system behaves as a paramagnet with effective magnetic moment that comes mostly from the Yb $4f$ electrons, although the somewhat enhanced values in comparison to the Yb$^{3+}$ effective moment (see Table~\ref{Suscept1}) points to additional magnetic contributions from the Fe and/or Zn atoms.\cite{Mardegan}
At low temperatures the screening of the magnetic moments by the $ce$ causes a suppression of the magnetic response, leaving a Pauli-like enhanced magnetic susceptibility at $\chi(T=0)$.
For comparison, elemental Pd is a transition metal with $d$-electrons that are on the verge of magnetism, and features $\chi(T=0)=0.75\times10^{-3}~emu/mol$,\cite{Stewart} which is still two orders of magnitude smaller than the obtained values summarized in Table~\ref{Suscept1} for these heavy fermion compounds. 
In this limit we have an enhanced magnetic susceptibility $\chi(T=0)$ that gives information about electron correlations in the system, when compared with the Sommerfeld coefficient obtained in the specific heat measurements.

With this in mind, we calculate the Wilson ratios $R_W=\chi/\gamma$ and the results (see Tab.~\ref{tab2}) show that Cd substitution slightly reduces this parameter from 1.3 in the ternary to 1.1 in the sample with the highest amount of Cd, in agreement with a reduction of the electron-electron correlation in the point of view of the Landau parameter.\cite{Stewart}
Thus, a shift in the direction from magnetic to non-magnetic response in the YbFe$_{2}$Zn$_{20-x}$Cd$_{x}$ is supported by the DC magnetic susceptibility measurements, whose maximum value also decreases as a function of Cd content.

We note that although there is a strong reduction in both $\chi(T=0)$ and $\gamma$ with the Cd substitution, there is only a slight decrease in the Wilson ratio. 
This means that it is possible to observe a reduction of the hybridization in both the specific heat and magnetic susceptibility measurements.
Theoretical investigations of the pressure effects on Yb based compounds\cite{Hai}, show that the linear coefficient of the specific heat and Pauli-like magnetic susceptibility increase with positive pressure, therefore consistent with the idea of negative pressure effects on Yb inside the Zn cage supported by our experimental results.

The inset of Fig.~\ref{Suscept} highlights a concomitant raise of the temperature associated with the broad maximum in magnetic susceptibility, from 12~K for the ternary to 17~K for the highest Cd-doped sample. 
Since this temperature of the broad maximum is associated with the Kondo temperature of the system, we can state $apriori$ that $T_K$ is increasing with the Cd doping (in agreement with the Doniach diagram for negative pressure in the Yb ion\cite{Doniach}) as will be shown ahead. 

The Kondo physics behind this varying hybridization system can also be discussed based on the transport measurements (Fig.~\ref{Res}). 
The Cd doping has a strong effect in the transport properties as can be seen in the reduction of the residual resistivity $\rho_{0}$, in the slope of the linear regime at low temperatures in the $\rho$~\emph{vs.}~$T^2$ plot (associated with the $A$ parameter, see inset) and also in the behavior at high temperatures.

\begin{figure}[!ht]
\begin{center}
\includegraphics[width=85mm,keepaspectratio]{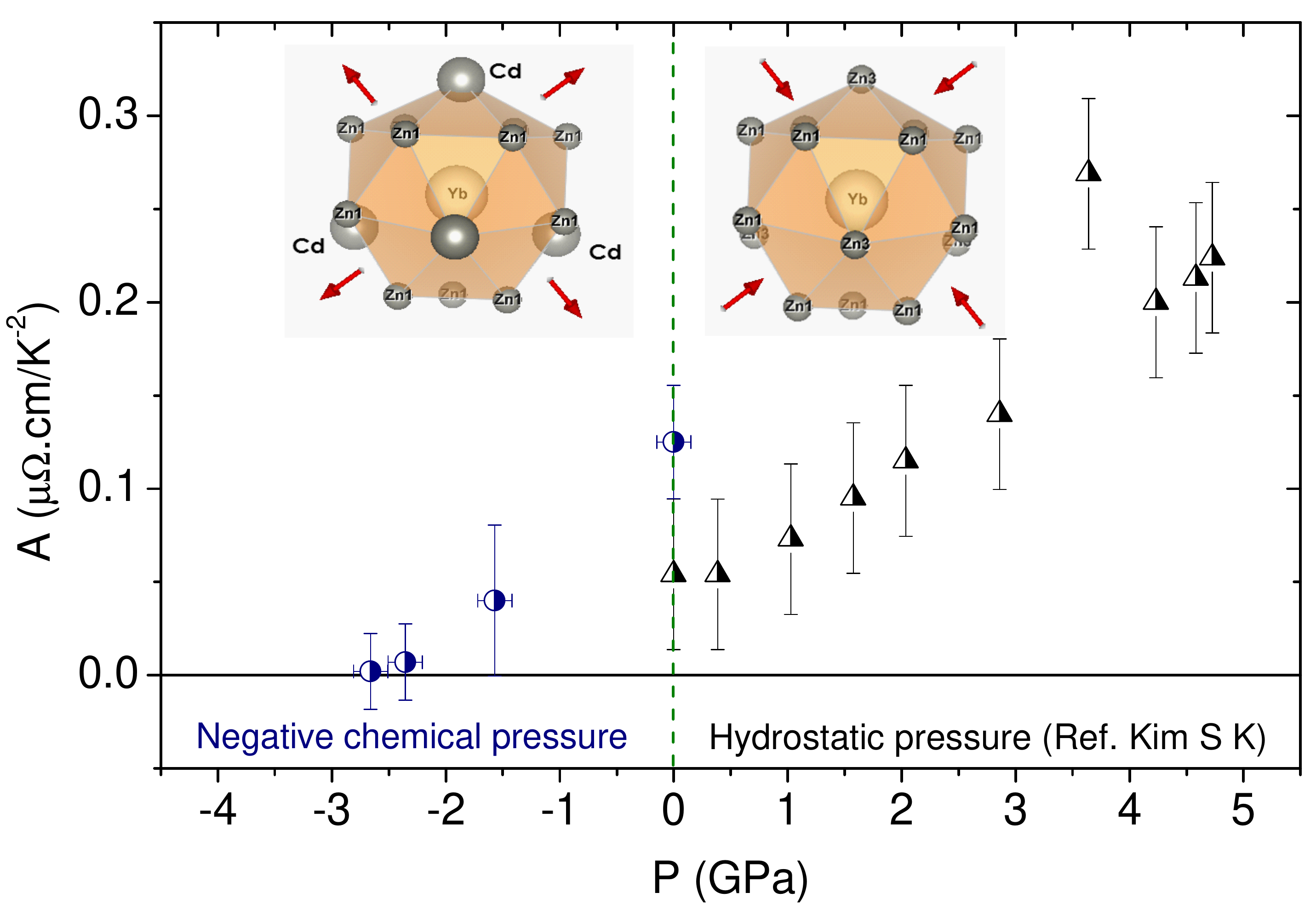}
\end{center}
\vspace{-0.5cm} \caption{Overall evolution of the Fermi-liquid $A$ parameter as a function of positive hydrostatic pressure and negative chemical pressure for YbFe$_{2}$Zn$_{20}$.}
\label{Aparameter}
\end{figure}

Reported data of hydrostatic pressure in YbFe$_{2}$Zn$_{20}$ showed that the residual resistivity increases as a function of pressure\cite{Kim} and is in agreement with the reduction of $\rho_{0}$ for negative pressures (see Tab.~\ref{tab2}).
The inset of Fig.~\ref{Res} shows that the $A$ parameter, which is proportional to the density of states at the Fermi level,\cite{Tsujii} is reduced with Cd substitution, in agreement with the heat capacity and dc magnetic susceptibility.

In order to follow the evolution of the $A$ parameter as a function of pressure, we must estimate the values of pressure that correspond to the expansion of the lattice parameter. 
This was achieved by adopting a representative value of 160 GPa for the bulk modulus of this family of compounds\cite{Chen} and using Murnaghan's equation of state.\cite{Murnaghan}
Fig.~\ref{Aparameter} shows the overall evolution of the $A$ parameter from high positive physical pressures (adapted from Ref.~\cite{Kim}) to our negative chemical pressure estimates.
There is a very good match of the two independent techniques, and the trend demonstrates a complete recovery of the Fermi Liquid behavior at negative pressures. 

Lastly, we address the broadening of the coherence shoulder at low temperatures ($T\leq25$~K) with the Cd substitution and a reduction of the slope of the linear regime at high temperatures in the main panel of Fig.~\ref{Res} ($\rho$~\emph{vs.}~$T$), associated with a general reduction of electron-phonon scattering in the entire temperature range.

\begin{figure}[!ht]
\begin{center}
\includegraphics[width=85mm,keepaspectratio]{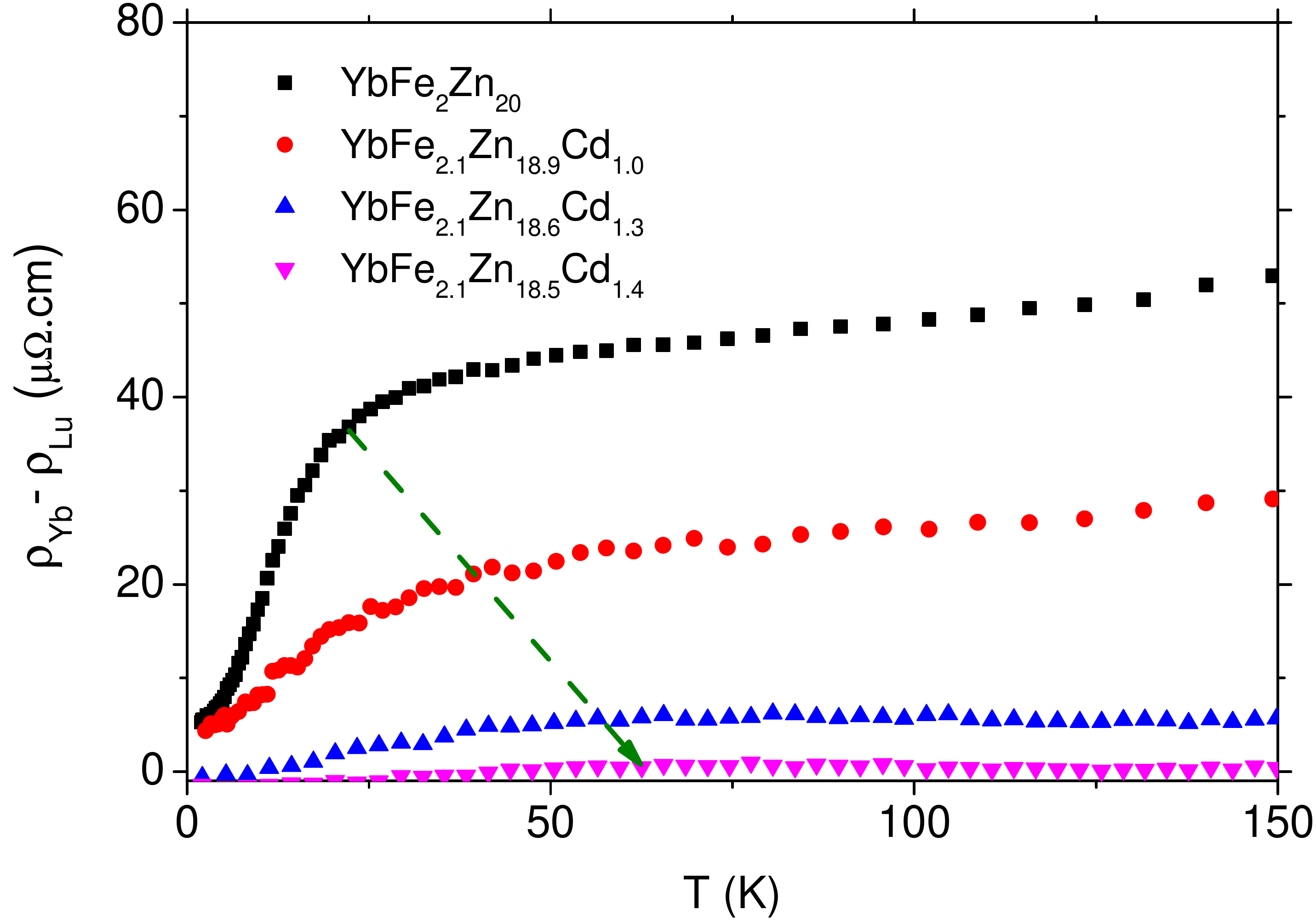}
\end{center}
\vspace{-0.5cm} \caption{Magnetic contribution to the electrical resistivity of YbFe$_{2}$Zn$_{20-x}$Cd$_{x}$ ($0 \leq x \leq 1.4$) obtained after subtracting the resistivity of the reference compound LuFe$_{2}$Zn$_{20}$.}
\label{Resre}
\end{figure}

The influence of the hybridized Yb ion in this unusual behavior can be better understood by comparing this magnetic system with its ``non-magnetic'' reference compound LuFe$_{2}$Zn$_{20}$ (here we are neglecting any minor magnetic scattering from Fe and Zn).
Fig.~\ref{Resre} shows the magnetic contribution to the resistivity ($\rho_{Yb}$ - $\rho_{Lu}$) of our YbFe$_{2}$Zn$_{20-x}$Cd$_{x}$ crystals, obtained by subtracting the resistivity of LuFe$_{2}$Zn$_{20}$ (Ref.~\cite{Kim}) from those in Fig.~\ref{Res}.
The undoped magnetic system rapidly loses the coherent peak as Cd is introduced and approaches the behavior of the reference compound.
As argued previously, the system is going in the direction of non-magnetic response, supporting the idea of a valence-shift of the Yb ion.

A final piece of valuable information can be extracted from these resistivity measurements. 
The Kadowaki-Woods Ratio\cite{Kadowaki} (KWR = $A$/$\gamma ^{2}$) for each Cd concentration was calculated and summarized in Table~\ref{tab3}.
Before the Kondo screening takes place, the ground state degeneracy of the Yb $4f$ levels due to the crystalline electric field splitting has an important role in the definition of the value of the Kondo temperature. 
In agreement with the generalized Kadowaki-Woods plot,\cite{Tsujii,Torikachvili} our obtained values suggest that in the ternary (YbFe$_{2}$Zn$_{20}$) the Yb ion has $N=6$ degeneracy of the ground state, but with Cd substitution the system goes toward $N=8$ degeneracy. 
With this information and using\cite{Torikachvili} $T_{K}=(RlnN)/\gamma$ we can estimate the evolution of the Kondo temperature, presented in Table~\ref{tab3}). 

\begin{table}
\centering
\caption{Cd concentrations ($x$), Kadowaki-Woods Ratio (KWR) and Kondo temperature (T$_{K}$) for the YbFe$_{2}$Zn$_{20-x}$Cd$_{x}$ system.}
\label{tab3}
\vspace{+0.4cm}
\begin{tabular}{|c||c||c|}
 \hline
 Conc.   &   $KWR$$\times$10$^{-7}$  & T$_{K}$\\
\hline
\emph{x}  &    ($\frac{\mu \Omega.cm.mol^{2}K^{2}}{mJ^{2}}$)  & K\\
  \hline
 0                                 &   4.74(2)  & 32(1)\\
 \hline
 1.0                            &   1.60(2)  & 35(1)\\
 \hline
 1.3                            &   0.39(2)  &40(1)\\
 \hline
 1.4                             &   0.15(2)  &41(1) \\   
 \hline 
\end{tabular}
\end{table}

As previously anticipated by the position of the maximum in DC magnetic susceptibility (inset Fig.~\ref{Suscept}), the Kondo temperature increases as a function of the Cd substitution. 
Once again, the Cd-doping effect is seen to move the system towards a non-magnetic state, supporting the idea of a valence shift towards divalent Yb.

\section{Conclusions}

An Yb valence shift due to the negative chemical pressure effect resultant from Cd-doping of the heavy fermion YbFe$_{2}$Zn$_{20}$ has been evidenced, supported by strong changes observed in XRD, heat capacity, magnetic susceptibility and resistivity measurements as a function of temperature on samples with increasing Cd content (up to 7\%).
The negative chemical pressure drives the Kondo temperature of the system to higher values, and the regime of Fermi-liquid behavior is extended, which agrees with and complements previous work applying  positive physical pressure. 
The results demonstrate a high efficiency of the Cd doping in this type of system, towards the tuning of the RKKY and the Kondo interaction strengths. 
We expect that this work can provide key reference elements to help understand the behaviors of related heavy fermions in the same family, such as YbCo$_{2}$Zn$_{20}$ in which much stronger $4f-ce$ hybridization is found.

\vspace{2pc}

\section{Acknowledgments}
We are very grateful to Complexo Laboratorial Nanotecnol\'{o}gico (CLN) - UFABC - SisNano for the XRD measurements, and to C. Rettori, P. G. Pagliuso and L. M. Ferreira for a critical reading of the manuscript.
This work was supported by Brazilian agencies FAPESP (Grant Nos. 2011/19924-2, 2012/17562-9), CNPq (Grant No. CNPq 402289/2013-7), FINEP and CAPES.

\end{document}